\documentclass[english]{iopart}
\usepackage[T1]{fontenc}
\usepackage[latin9]{inputenc}

\usepackage{iopams}
\usepackage{setstack}

\usepackage{babel}

\begin{document}

\title{Spin and angular momentum operators and their conservation}

\author{Michael Mazilu}

\address{SUPA, School of Physics and Astronomy, University of St Andrews,
St Andrews KY16 9SS, UK}

\ead{mm17@st-andrews.ac.uk}
\begin{abstract}
Lorentz's reciprocity lemma and Feld-Tai reciprocity theorem show the effect of interchanging the action and reaction in Maxwell's equations. We derive a free-space version of these reciprocity relations which generalizes the conservation of the momentum-energy tensor. This relation corresponds to the interference conservation of electromagnetic waves. We show that for any transformation or symmetry that leaves Maxwell's equations invariant, we can modify the reciprocity relation to introduce a conserving density, optical flux and stress tensor extending Noether's theorem to a different context. We apply this method to transformations that can be expressed as Hermitian operators and more specifically, we define the operators associated with the optical energy, spin, linear and angular momentum.
\end{abstract}

\maketitle

\section{Introduction}

Defining the optical spin and angular momentum density of the electromagnetic field is an important aspect in the study of optical vortices and their evolution and dynamics. Indeed, optical vortices are associated with a circular flow of energy, but how can we define a physically meaningful quantity that helps us understand this flow? One possible approach is to use operators taken from Quantum Mechanics on the electromagnetic fields. This can be done within the paraxial approximation \cite{VANENK:1992p9091,Berry:1998p9093} or based on the second quantization \cite{VanEnk:1994p9089}. Within this paraxial approximation, Maxwell's equations become scalar and are formally identical to Schr\"odinger's equations making it possible to use quantum mechanical operators.  Unfortunately, defining such quantities for vectorial fields is not as straight forward. Various similar methods have been proposed in the literature based on the angular momentum of the Poynting vector \cite{Barnett:1994p9050,Barnett:2002p9094}. More recently, the energy flow around vortices \cite{Bekshaev:2007p9088} has been equated to optical currents  by invoking the "democratic" principle between the electric and magnetic fields \cite{Berry:2009p9030}. It is this last approach that we build upon and generalise in this paper by showing crucially that this optical current is only one part of whole conserving energy-momentum tensor. The method presented here is based on the electromagnetic reciprocity theorem and the superposition principle.  The generality of this approach makes it applicable to other conserving currents, each corresponding to a symmetry, and even to other field evolution equations.  Finally, we conclude by discussing the link between these conserving currents, the symmetry operators and the quantum mechanical operators.

\section{Free space reciprocity relations}
The starting point of our paper is the optical reciprocity theorem, which defines Newton's third law for optical fields \cite{Potton:2004p8923,Cornille:1998p8987}. More generally, the reciprocity relations describe what happens when cause and effect are exchanged. For Maxwell's equations, the reciprocity theorem takes two forms. The Lorentz reciprocity lemma which links the electric field created by an emitting current and the electric field from the receiving current and the Feld-Tai reciprocity theorem which links the same currents but via the generated magnetic fields \cite{Tai:1992p8927,Feld:1992p9096}. Here, we are not interested in the current-field interaction part of the reciprocity relation but only in the electromagnetic free-field behaviour within the reciprocity theorem. In this case, the reciprocity theorems simplify to an interference conserving relation linking two independent solution of Maxwell's equations. In the appendix, we derive this reciprocity relation in the case of the Riemann-Silberstein form of Maxwell's equations \cite{Berry:2004p8934,BialynickiBirula:2003p8941} that formally is similar to Schr\"odinger's equations. 

Let us consider Maxwell's equations in free space,
\begin{equation}\label{Eq:Maxwell}\eqalign{
\nabla\cdot\epsilon_{0}\mathbf{E}  =  0,\\
\nabla\cdot\mu_{0}\mathbf{H}  =  0,\\
\nabla\times\mathbf{E}  =  -\mu_{0}\partial_{t}\mathbf{H},\\
\nabla\times\mathbf{H}  =  \epsilon_{0}\partial_{t}\mathbf{E},
}\end{equation}
where $\mathbf{E}$ and $\mathbf{H}$ are the complex electric and magnetic 
vector fields  and where $\epsilon_{0}$ and $\mu_{0}$ the vacuum permittivity and permeability. The energy and momentum associated with these fields is given by ${E} $ the energy density, ${\bf S}$ the Poynting vector and $\widetilde{\sigma}$ Maxwell's stress tensor \cite{Bateman:1922p210} 
\begin{equation}\eqalign{
E(\mathcal{F})  = & \frac{1}{2}\left(\epsilon_{0}{\bf E}^*\cdot\mathbf{E}+\mu_{0}{\bf H}^*\cdot{\bf H}\right),\cr
{\bf S}(\mathcal{F})  =&\frac{1}{2}\left(  {\bf E}^*\times{\bf H} + {\bf E}\times{\bf H}^*\right),\cr
\widetilde{\sigma}(\mathcal{F})  =&\frac{c^2}{2}\left( \left(\epsilon_{0}{\bf E}^*\cdot\mathbf{E}+\mu_{0}{\bf H}^*\cdot{\bf H}\right)\widetilde{I}- \epsilon_{0}{\bf E}^*\otimes{\bf  E} \right. \\
& \left. -\epsilon_{0}{\bf E}\otimes{\bf  E}^*-\mu_{0}\mathbf{H}^*\otimes\mathbf{H}-\mu_{0}\mathbf{H}\otimes\mathbf{H}^*\right),}
\end{equation}
 where $\otimes$ stands for the tensor product, and $*$ stands for the complex conjugate,  $c=1/\sqrt{\epsilon_{0}\mu_{0}}$ for the speed of light, $\widetilde{I}$
for the identity 3x3 matrix and $\mathcal{F}$ for the pair of vectors $\mathcal{F}=(\mathbf{E},\mathbf{H})$. The conservation of energy and momentum of the electromagnetic field is given by: 
\begin{equation}\label{Eq:classical-conservation}\eqalign{
\nabla\cdot{\bf S}(\mathcal{F})+\partial_{t}{E}(\mathcal{F})  =  0,\cr
\nabla\cdot\widetilde{\sigma}(\mathcal{F})+\partial_{t}{\bf S}(\mathcal{F}) =  0.
}\end{equation}

An intuitive way to deduce the free space reciprocity relations is through the use of linear superposition \cite{Cornille:1998p8987}.  Here, we have simplified this approach by considering only the linear superposition $\mathcal{F}= \mathcal{F}_{1}+ \mathcal{F}_{2}$
of two fields $\mathcal{F}_{1}$ and $\mathcal{F}_{2}$,
both solutions of Maxwell's equations (\ref{Eq:Maxwell}). As the energy, momentum and stress tensor are all quadratic in the field, each of them can be split these into three parts. Two parts involve only terms in $\mathcal{F}_1$ and $\mathcal{F}_{2}$ respectively while the remaining terms involve cross products between these fields. This decomposition can the represented as: 
\begin{equation}\label{Eq:ESdecomp}\eqalign{
{E} (\mathcal{F}) =  {E}(\mathcal{F}_{1})+{E}_{12}(\mathcal{F}_1, \mathcal{F}_2)+{E}(\mathcal{F}_{2}),\cr
{\bf S} (\mathcal{F}) =  {\bf S}(\mathcal{F}_{1})+{\bf S}_{12}(\mathcal{F}_1, \mathcal{F}_2)+{\bf S}(\mathcal{F}_{2}),\cr
\widetilde{\sigma} (\mathcal{F}) =  \widetilde{\sigma}(\mathcal{F}_1)+\widetilde{\sigma}_{12}(\mathcal{F}_1, \mathcal{F}_2)+\widetilde{\sigma}(\mathcal{F}_{2}),
}\end{equation}
where the expression with the subscript $_{12}$ groups all the terms involving cross products between the two fields. These expression are given by:
\begin{equation} \label{EQ:ESs12}\eqalign{
{E}_{12} (\mathcal{F}_1, \mathcal{F}_2) = &  
\frac{1}{2}\left(\epsilon_{0}\mathbf{E}^*_{1}\cdot\mathbf{E}_{2}+\mu_{0}\mathbf{H}^*_{1}\cdot\mathbf{H}_{2}\right),\\
{\bf S}_{12}(\mathcal{F}_1, \mathcal{F}_2)  = & 
\frac{1}{2}\left( \mathbf{E}^*_{1}\times\mathbf{H}_{2}+\mathbf{E}_{2}\times\mathbf{H}^*_{1}\right),\\
\widetilde{\sigma}_{12}(\mathcal{F}_1, \mathcal{F}_2)  = &  
\frac{c^2}{2}\left(\left(\epsilon_{0}\mathbf{E}^*_{1}\cdot\mathbf{E}_{2}+\mu_{0}\mathbf{H}^*_{1}\cdot\mathbf{H}_{2}\right) \widetilde{{I}}-\epsilon_{0}\mathbf{E}^*_{1}\otimes\mathbf{E}_{2}\right. \\
& \left. -\epsilon_{0}\mathbf{E}_{2}\otimes\mathbf{E}^*_{1}  -\mu_{0}\mathbf{H}^*_{1}\otimes\mathbf{H}_{2}-\mu_{0}\mathbf{H}_{2}\otimes\mathbf{H}^*_{1}\right).}
\end{equation}
These expressions describe the energy and momentum associated with the interference of the two fields and correspond to its conservation. Indeed, each of the two fields $\mathcal{F}_{1}$ and $\mathcal{F}_{2}$ fulfill the conservation relations (\ref{Eq:classical-conservation}). Therefore, when substituting  (\ref{Eq:ESdecomp}) into (\ref{Eq:classical-conservation}) we find: 
\begin{equation} \label{EQ:consEH}\eqalign{
\nabla\cdot{\bf S}_{12}(\mathcal{F}_1, \mathcal{F}_2)+\partial_{t}{E}_{12}(\mathcal{F}_1, \mathcal{F}_2)  =  0,\cr
\nabla\cdot\widetilde{\sigma}_{12}(\mathcal{F}_1, \mathcal{F}_2)+\partial_{t}{\bf S}_{12}(\mathcal{F}_1, \mathcal{F}_2)  =  0.}
\end{equation}
These relations describe the conservation  of the interference of two solution of Maxwell's equations and represent the free-space equivalent to Lorentz lemma \cite{Potton:2004p8923} and the Feld-Tai reciprocity relations \cite{Tai:1992p8927,Feld:1992p9096}. Physically, relation (\ref{EQ:consEH}) means that if the superposition of two fields is associated with a specific interference energy density,  then energy density this will be conserved regardless of the linear propagation transformation acting on each of the fields. Otherwise stated, it is not possible to treat the two field distributions independently of each other which is equivalent to a classical energy entanglement (property that has also been observed in \cite{Basini:2003p185}). In the appendix, we present the same procedure for the Riemann-Silberstein form of Maxwell's equations. 

\section{Invariant transformations and symmetries}

In this section, we consider the effect of a field or space transformation, which leaves Maxwell's equations invariant, on the free space reciprocity relation defined in the second section. These transformations are each linked to a symmetry of Maxwell's equations and we show that for each such symmetry there exists a conserving optical current or more generally a canonical conserving energy-momentum tensor. Our result extends Noether's theorem \cite{Kaku:1994p9111} to the reciprocity theorem offering a direct way to deduce symmetric canonical energy-mometum tensors from each other (see  \cite{Munoz:1996p2990} for a discussion of the symmetry problem).  This first result of the paper is stated in the form of a theorem:

{\em If $\mathcal{F}$ is a solution of Maxwell's equations, $\mathbb{T}$ a transformation that leaves Maxwell's equations invariant and ${\mathbb{T}} \mathcal{F}$ is the  transformed field still solution of Maxwell's equation, then ${E}_{12}\left(\mathcal{F},\mathbb{T} \mathcal{F}\right)$, $\mathbf{S}_{12}\left(\mathcal{F}, \mathbb{T} \mathcal{F}\right)$ and $\widetilde{\sigma}_{12}\left(\mathcal{F},\mathbb{T} \mathcal{F}\right)$ define the canonical energy density, current and stress tensor associated with this transformation or symmetry. These quantities form the canonical energy-momentum tensor and verify the conservation relation (\ref{EQ:consEH}). } 

This theorem can be proved directly by substituting $\mathcal{F}_1= \mathcal{F}$ and $\mathcal{F}_2=\mathbb{T} \mathcal{F}$ in the reciprocity relations (\ref{EQ:ESs12}) and (\ref{EQ:consEH}). This means that for every invariant transformation of  Maxwell's equations, we can construct a conserving quantity together with its conserving current. This result is related to Noether's theorem that uses invariant symmetries of the Lagrangian to deduce, for each symmetry, a canonically conserving current \cite{Kaku:1994p9111}.

Additionally for each symmetry or invariant transformation, we can define the following primary integral of the system:
\begin{equation}\eqalign{
<{E}_{12}(\mathcal{F})>_{\mathbb T}&=\int  {E}_{12}\left(\mathcal{F},{\mathbb T} \mathcal{F}\right)\mathrm{d^3}r,\cr
<\mathbf{S}_{12}(\mathcal{F})>_{\mathbb T}&=\int  \mathbf{S}_{12}\left(\mathcal{F},{\mathbb T} \mathcal{F}\right)\mathrm{d^3}r.
}\end{equation}
These conserving canonical quantities can be associated with each solution and symmetry of Maxwell's equation and are constant in time. Here, we note that these quantities  are not linear with the respect to the field but quadratic. A prime example of such a conserving quantity is the energy of the electromagnetic wave which is $<E_{12}(\mathcal{F})>_{\mathbb I}$ where the transformation corresponds to the identity transformation $\mathbb I$. 

\section{Canonical superposition principle }

In the fourth section, we deal with the interplay between conservation, superposition and interference of independent solutions of Maxwell's equations. Indeed, a superposition of multiple such solutions is still a solution. However, because of the interference between these different solutions, the conserving quantities are not necessarily additive. Starting from this observation, we formulate the principle of canonical superposition implying a simultaneous linearity in the fields and in the conserving quantities in a linear superposition. To satisfy this principle, the electromagnetic field needs to be decomposed onto the eigenfunctions of the operators associated with each symmetry. Additionally, for Hermitian operators we show that a complete sets of pair-wise commuting operators correspond to compatible conserving quantities. However, there are no electromagnetic fields that simultaneously satisfy the canonical superposition principle for non-commuting operators. This approach enables the introduction of the photon wave function \cite{BialynickiBirula:2006p8992} and a first quantisation of Maxwell's equation \cite{delaTorre:2005p9072}. 

In this section, we restrict ourselves to invariant transformations that are linear and can be expressed as a linear operator $\mathbb A$ acting on the complex  vector field  $\mathcal{F}$ defined earlier. The second result of this paper can be written in the form of the principle: 

\emph{For a vector field to be observed as an independent entity when interfering with another external field, its conserving canonical quantities must be additive to those  of the external field.}

Mathematically this can be expressed in the following way. Let the total field $\mathcal{F}$ be decomposed into a superposition of fields $\mathcal{F}=\sum_i \mathcal{F}_i$ where each $\mathcal{F}_i$ is a solution of Maxwell's equations. For each part of the field $\mathcal{F}_i$ to be observed as an independent entity, the total conserving canonical quantity must be a sum of all the individual conserving canonical quantities
\begin{equation} \eqalign{
<E_{12}(\mathcal{F})>_{{\mathbb A}}&=\left<E_{12}\left({\sum}_i \mathcal{F}_i\right)\right>_{{\mathbb A}},\cr
&=\sum_i <E_{12}({\mathcal{F}_i})>_{{\mathbb A}}.
}\end{equation}

Considering linear Hermitian operators ${\mathbb A}$ that leave Maxwell's equations invariant, we can verify that their eigenfunctions are fulfilling the canonical superposition principle. Indeed, we have:
\begin{equation} \eqalign{
<E_{12}(\mathcal{F})>_{{\mathbb A}}&=\sum_i \lambda_i <E_{12}(\mathbf{\mathcal{F}_i})>_{{\mathbb A}}
}\end{equation}
where $\lambda_i$ is the eigenvalue of $\lambda_i \mathcal{F}_i={\mathbb A} \mathcal{F}_i$. This is due to the fact that the eigenfunctions of a Hermitian operator are orthogonal  and as such their interference integrated over the whole space cancels out. 

To appreciate that the canonical superposition principle is not automatically satisfied, we can look at the example of adding two identical electromagnetic fields similar to the process of stimulated emission. 
This superposition increases the total energy by a factor of 4. For the canonical superposition principle to be fulfilled with two identical fields, the two fields need to be in phase quadrature with respect to each other. In this case, adding the two fields together also adds their total energy. This means that the field associated with the stimulated emission needs to be out of phase with respect to the incident field such that its energy and field distribution are adding correctly.

Finally, considering the simultaneous canonical superposition principle of multiple conserving quantities, we remark that there are quantities that are mutually exclusive. Indeed, for the superposition to work simultaneously for multiple linear operators, the eigenfunctions of these operators need to be shared. This is only possible for an ensemble of Hermitian operators that commute pair-wise.  The maximum number of quantities that fulfil the canonical superposition principle is given by a complete set of commuting operators. This also implies that there are conserving quantities that are mutually exclusive i.e. it is not possible to define a solution of Maxwell's equation that, at the same time, can be treated as an entity, conserving in a superposition, these canonical quantities.

\section{Applications}

In the final section, we apply the canonical superposition principle to define  the spin, dynamic energy,  linear and angular momentum currents. Four operators associated with these quantities form a complete set of pair-wise commuting operators with the non-diffracting Bessel beam as a common eigenfunction. We also verify the ratio between the dynamic energy and spin to be equal to the optical frequency \cite{OHANIAN:1986p9042}. The form of the spin current and density derived using this method is proportional to that found in the literature \cite{Barnett:2002p9094,Berry:2009p9030}.

\subsection{Classical energy-momentum}

Here, we remark that deriving the classical energy-momentum tensor from the free space reciprocity relations is straightforward. The transformation needed in this case is the identity transformation and implies the following relations:
\begin{equation}\label{Eq:classical-energy}\eqalign{
{\mathbb A}= \mathbb{I}, \\
{E}_{12}\left(\mathcal{F},{\mathbb A} \mathcal{F}\right)  = & \frac{1}{2}\left(\epsilon_{0}{\bf E}^*\cdot\mathbf{E}+\mu_{0}{\bf H}^*\cdot{\bf H}\right),\cr
{\bf S}_{12}\left(\mathcal{F},{\mathbb A} \mathcal{F}\right)   = &\frac{1}{2}\left(  {\bf E}^*\times{\bf H} + {\bf E}\times{\bf H}^*\right),\cr
\widetilde{{\bf \sigma}}_{12}\left(\mathcal{F},{\mathbb A} \mathcal{F}\right) =&\frac{c^2}{2}\left( \left(\epsilon_{0}{\bf E}^*\cdot\mathbf{E}+\mu_{0}{\bf H}^*\cdot{\bf H}\right)\widetilde{{I}}- \epsilon_{0}{\bf E}^*\otimes{\bf  E} \right. \\
& \left. -\epsilon_{0}{\bf E}\otimes{\bf  E}^*-\mu_{0}\mathbf{H}^*\otimes\mathbf{H}-\mu_{0}\mathbf{H}\otimes\mathbf{H}^*\right).}
\end{equation}

\subsection{Dynamic energy-momentum }

The first non-trivial Hermitian operator that leaves Maxwell's equations invariant is the derivative with respect to the time co-ordinate. In Quantum Field theory the time derivative operator corresponds to the invariance to the time translation and is associated with the energy of the system \cite{Kaku:1994p9111}. Using this derivative, all the static parts of the electromagnetic fields are eliminated and only the dynamic parts remains. The imaginary coefficient is necessary to maintain hermiticity of the dynamic operator and the energy-momentum tensor is given by:
\begin{equation}\label{Eq:w-energy}\eqalign{
{\mathbb A}_t=i\partial_t, \\
{E}_{12}\left(\mathcal{F},{\mathbb A}_t \mathcal{F}\right)  = & \frac{i}{2}\left(\epsilon_{0}{\bf E}^*\cdot\partial_t\mathbf{E}+\mu_{0}{\bf H}^*\cdot\partial_t{\bf H}\right),\cr
{\bf S}_{12}\left(\mathcal{F},{\mathbb A}_t \mathcal{F}\right)   = &\frac{i}{2}\left(  {\bf E}^*\times\partial_t{\bf H} + \partial_t{\bf E}\times{\bf H}^*\right),\cr
\widetilde{{\bf \sigma}}_{12}\left(\mathcal{F},{\mathbb A}_t \mathcal{F}\right) =&\frac{ic^2}{2}\left( \left(\epsilon_{0}{\bf E}^*\cdot\partial_t\mathbf{E}+\mu_{0}{\bf H}^*\cdot\partial_t{\bf H}\right)\widetilde{{I}}- \epsilon_{0}{\bf E}^*\otimes\partial_t{\bf  E} \right. \\
& \left. -\epsilon_{0}\partial_t{\bf E}\otimes{\bf  E}^*-\mu_{0}\mathbf{H}^*\otimes\partial_t\mathbf{H}-\mu_{0}\partial_t\mathbf{H}\otimes\mathbf{H}^*\right).}
\end{equation}

We remark that the eigenfunctions of the dynamic operator $i\partial_t$ are monochromatic waves with an optical frequency defined as $\omega$. For these waves, the dynamic energy-momentum tensor is proportional to the classical energy (\ref{Eq:classical-energy}) and to $\omega$. Unit-wise, the energy units can be recovered by making the electric and magnetic fields unit-less and using $\hbar$ as the conversion factor. Finally, using this conversion factor we can also introduce the momentum operator ${\mathbb A}_p=i\hbar\nabla$ which corresponds in Quantum Field theory to the translation independence.  

\subsection{Optical spin flux}

The optical spin is associated with the circular polarisation of the light field. Here, we look for an Hermitian operator whose eigenfunctions are circular polarised light fields, which can be written as 
$${\mathbb A}_s=\left(\begin{array}{cc} 0& iZ_0\\-i/ Z_0& 0 \end{array}\right) $$
where $Z_0=\mu_0 c$ is the vacuum impedance. This operator corresponds to the duality transformation which leaves Maxwell's equations invariant. We also remark that the vacuum impedance does not break the Hermiticity of this operator as its purpose is to convert between the electric and magnetic field. The corresponding energy-momentum tensor is given by:
\begin{equation}\label{Eq:spin}\eqalign{
{E}_{12}\left(\mathcal{F},{\mathbb A}_s \mathcal{F}\right)  = & \frac{i}{2c}\left({\bf E}^*\cdot\mathbf{H}-{\bf H}^*\cdot{\bf E}\right),\cr
{\bf S}_{12}\left(\mathcal{F},{\mathbb A}_s \mathcal{F}\right)   = &\frac{ic}{2}\left( \epsilon_{0} {\bf E}^*\times{\bf E} +\mu_{0} {\bf H}\times{\bf H}^*\right),\cr
\widetilde{{\bf \sigma}}_{12}\left(\mathcal{F},{\mathbb A}_s \mathcal{F}\right) =&\frac{ic}{2}\left( \left({\bf E}^*\cdot\mathbf{H}-{\bf H}^*\cdot{\bf E}\right)\widetilde{{I}}- {\bf E}^*\otimes{\bf  H} \right. \\
& \left. -{\bf H}\otimes{\bf  E}^*+\mathbf{H}^*\otimes\mathbf{E}+\mathbf{E}\otimes\mathbf{H}^*\right).}
\end{equation}

This describes the conservation of the circular polarisation as the light field propagates. Indeed, all the terms in the optical spin energy-momentum tensor are zero for linearly polarised light. Additionally, a circularly polarised plane wave has a constant optical spin density and the sign of this density depends on the polarisation handedness. The conservation of this definition of the optical spin has been theoretically asserted 
\cite{Barnett:1994p9050,Barnett:2002p9094,Berry:2009p9030}  but not its associated stress tensor meaning it is not Lorentz invariant. Finally, the ratio between the optical spin density and the dynamic energy density as defined by (\ref{Eq:w-energy}) is, as expected, $\omega$ \cite{OHANIAN:1986p9042}.

\subsection{Optical angular momentum}

The angular momentum operator takes advantage of the cylindrical symmetry. Here, we have expressed it in cylindrical co-ordinates and we consider only a momentum in the $z$ direction where $\phi$ is the azimuthal angle:
\begin{equation}\label{Eq:orbit-energy}\eqalign{
{\mathbb A}_\phi=i\partial_\phi, \\
{E}_{12}\left(\mathcal{F},{\mathbb A}_\phi \mathcal{F}\right)  = & \frac{i}{2}\left(\epsilon_{0}{\bf E}^*\cdot\partial_ \phi\mathbf{E}+\mu_{0}{\bf H}^*\cdot\partial_ \phi{\bf H}\right),\cr
{\bf S}_{12}\left(\mathcal{F},{\mathbb A}_\phi \mathcal{F}\right)   = &\frac{i}{2}\left(  {\bf E}^*\times\partial_ \phi{\bf H} + \partial_ \phi{\bf E}\times{\bf H}^*\right),\cr
\widetilde{{\bf \sigma}}_{12}\left(\mathcal{F},{\mathbb A}_\phi \mathcal{F}\right) =&\frac{ic^2}{2}\left( \left(\epsilon_{0}{\bf E}^*\cdot\partial_ \phi\mathbf{E}+\mu_{0}{\bf H}^*\cdot\partial_ \phi{\bf H}\right)\widetilde{{I}}- \epsilon_{0}{\bf E}^*\otimes\partial_ \phi{\bf  E} \right. \\
& \left. -\epsilon_{0}\partial_ \phi{\bf E}\otimes{\bf  E}^*-\mu_{0}\mathbf{H}^*\otimes\partial_ \phi\mathbf{H}-\mu_{0}\partial_ \phi\mathbf{H}\otimes\mathbf{H}^*\right).}
\end{equation}
Higher order Laguerre-Gaussian and Bessel modes having a phase-front proportional to $\exp(il\phi)$ are eigenfunctions of this operator and their conserving canonical quantity corresponds to their azimuthal number $l$. Here again, making the electromagnetic field dimensionless and using Planck's constant $\hbar$ gives us the correct units for the angular momentum density. Additionally, the full momentum operator can be obtained by using ${\mathbb A}_L=i\hbar r\times\nabla$ which corresponds to a condensation of three operators that are non commutating between them. 

An interesting decomposition of the electromagnetic field can be achieved when considering the last three canonical conserving quantities. Indeed, the $z$-directional angular momentum operator ${\mathbb A}_\phi$ together with the dynamic energy ${\mathbb A}_t$, the $z$-momentum ${\mathbb A}_z$ and the spin ${\mathbb A}_s$ operators commute pair-wise and define a complete base of eigenfields. These vectorial electromagnetic fields correspond to circular polarized Bessel modes indexed using their azimuthal order, circular polarisation, longitudinal wave vector and optical frequency.  

\section{Conclusion}

By using the free-space reciprocity relation, we have introduced a method to generate conserving relations for every invariant transformation of Maxwell's equations. For Hermitian operators, we have shown that their eigenfunctions decompose any electromagnetic field into a superposition, where each part can be distinguished by its canonical conserving quantities. The choice of these quantities is restricted by the number of Hermitian operators that commute pair-wise. Finally, we applied this approach to define the dynamic energy momentum, the spin and angular momentum of the electromagnetic field. This results makes a direct first quantisation of Maxwell's equations possible and brings the concept of the classical electromagnetic photon wave function closer. Finally, by analogy it gives an insight into the real and imaginary part of the complex wave function in Quantum Mechanics (see Appendix). 

\section*{Acknowledgements}

I would like to thank Kishan Dholakia, Rab Wilson,  Aly Gillies and Petri\c sor Mazilu for valuable discussions, comments and for reading my article.  

\section*{Appendix: Riemann-Silberstein equations and their reciprocity relation}

In vacuum, Maxwell's equations (\ref{Eq:Maxwell}) are equivalent to:
\begin{equation}\label{eq:RS}\eqalign{
\nabla\cdot\mathbf{F}  =  0,\cr
\nabla\times\mathbf{F}  =  \frac{i}{c}\partial_{t}\mathbf{F},
}\end{equation}
where the complex vector field is defined as $ \mathbf{F}=(\mathbf{\sqrt{\epsilon_{0}}}\mathbf{E}+i\mathbf{\sqrt{\mu_{0}}}\mathbf{H})/2$ and where electric and magnetic fields considered here are purely real. These are the Riemann-Silberstein form of Maxwell's equations\cite{Berry:2004p8934,BialynickiBirula:2003p8941}. We can determine the reciprocity theorem for these equations by considering the interference currents and density for a superposition of two fields $\mathbf{F}=\mathbf{F}_{1}+\mathbf{F}_{2}$,
\begin{equation}\label{Eq:reciprocity_e}\fl\eqalign{
{E}_{RS}\left(\mathbf{F}_{1},\mathbf{F}_{2}\right)  =  \mathbf{F}_{1}^{*}\cdot\mathbf{F}_{2}+\mathbf{F}_{1}\cdot\mathbf{F}^*_{2},\cr
{\bf S}_{RS}\left(\mathbf{F}_{1},\mathbf{F}_{2}\right)  =  -ic(\mathbf{F}_{1}^{*}\times\mathbf{F}_{2}+\mathbf{F}^*_{2}\times\mathbf{F}_{1})\cr
\widetilde{{\bf \sigma}}_{RS}\left(\mathbf{F}_{1},\mathbf{F}_{2}\right)  =  -c^{2}(\mathbf{F}_{1}^{*}\otimes\mathbf{F}_{2}+\mathbf{F}_{2}\otimes\mathbf{F}_{1}^{*}+\mathbf{F}_{1}\otimes\mathbf{F}^*_{2}+\mathbf{F}^*_{2}\otimes\mathbf{F}_{1}-{E}\widetilde{{I}}),
}\end{equation}
where $\mathbf{F}_{1}$ and $\mathbf{F}_{2}$ are each solutions of equation (\ref{eq:RS}). Choosing $\mathbf{F}_{1}=\mathbf{F}_{2}$ in these definitions we find the energy, the flow and stress tensor associated with equations~(\ref{eq:RS}). 

The two conservation relations of the interference currents become then the reciprocity relations for two solutions $\mathbf{F}_{1}$ and $\mathbf{F}_{2}$ of (\ref{eq:RS}):
\begin{equation}\label{EQ:interference-conserv}\eqalign{
\nabla\cdot{\bf S}_{RS}\left(\mathbf{F}_{1},\mathbf{F}_{2}\right)+\partial_{t}{E}_{RS}\left(\mathbf{F}_{1},\mathbf{F}_{2}\right)  =  0,\cr
\nabla\cdot\widetilde{\sigma}_{RS}\left(\mathbf{F}_{1},\mathbf{F}_{2}\right)+\partial_{t}{\bf S}_{RS}\left(\mathbf{F}_{1},\mathbf{F}_{2}\right)  =  0.}
\end{equation}

\section*{Bibliography}

%\bibliographystyle{unsrt} 
%\bibliography{gll2}

\end{document}